\documentclass[reprint,amssymb,amsmath,aps,superscriptaddress,showpacs,pra,footinbib,longbibliography]{revtex4-1}
\usepackage{graphicx}
\usepackage{epstopdf}

\begin{document}

\title{Antiferroelectric instability in kagome francisites Cu$_3$Bi(SeO$_3$)$_2$O$_2$X (X = Cl, Br)}

\author{Danil A. Prishchenko}
\affiliation{Ural Federal University, Mira Str. 19, 620002 Ekaterinburg, Russia}

\author{Alexander A. Tsirlin}
\email{altsirlin@gmail.com}
\affiliation{Ural Federal University, Mira Str. 19, 620002 Ekaterinburg, Russia}
\affiliation{Experimental Physics VI, Center for Electronic Correlations and Magnetism, Institute of Physics, University of Augsburg, 86135 Augsburg, Germany}

\author{Vladimir~Tsurkan}
\affiliation{Experimental Physics V, Center for Electronic Correlations and Magnetism, Institute of Physics, University of Augsburg, 86135 Augsburg, Germany}
\affiliation{Institute of Applied Physics, Academy of Sciences Moldova, Chisinau MD-2028, Republic of Moldova}

\author{Alois Loidl}
\affiliation{Experimental Physics V, Center for Electronic Correlations and Magnetism, Institute of Physics, University of Augsburg, 86135 Augsburg, Germany}

\author{Anton Jesche}
\affiliation{Experimental Physics VI, Center for Electronic Correlations and Magnetism, Institute of Physics, University of Augsburg, 86135 Augsburg, Germany}

\author{Vladimir G. Mazurenko}
\affiliation{Ural Federal University, Mira Str. 19, 620002 Ekaterinburg, Russia}

\begin{abstract}
Density-functional calculations of lattice dynamics and high-resolution synchrotron powder diffraction uncover antiferroelectric distortion in the kagome francisite Cu$_3$Bi(SeO$_3$)$_2$O$_2$Cl below 115\,K. Its Br-containing analogue is stable in the room-temperature crystal structure down to at least 10\,K, although the Br compound is on the verge of a similar antiferroelectric instability and reveals local displacements of Cu and Br atoms. The I-containing compound is stable in its room-temperature structure according to density-functional calculations. We show that the distortion involves cooperative displacements of Cu and Cl atoms, and originates from the optimization of interatomic distances for weakly bonded halogen atoms. The distortion introduces a tangible deformation of the kagome spin lattice and may be responsible for the reduced net magnetization of the Cl compound compared to the Br one. The polar structure of Cu$_3$Bi(SeO$_3$)$_2$O$_2$Cl is only slightly higher in energy than the non-polar antiferroelectric structure, but no convincing evidence of its formation could be obtained.
\end{abstract}


\maketitle

\section{Introduction}
Novel magnetic materials with frustrated geometries attract attention of both theoretical and experimental communities~\cite{balents2010}. Kagome geometry of corner-sharing spin triangles hosts a number of interesting properties. Competing nearest-neighbor antiferromagnetic (AFM) exchange interactions lead to an infinitely degenerate classical ground state that gives way to a quantum spin liquid when quantum fluctuations are included~\mbox{\cite{yan2011,[{}][{, and references therein}]depenbrock2012}}. Spin-liquid ground state was experimentally observed in several spin-$\frac12$ Cu$^{2+}$ minerals, including herbertsmithite~\cite{mendels2010,han2012,fu2015} and kapellasite~\cite{fak2012}. A kagome-like geometry is also found in the mineral francisite Cu$_3$Bi(SeO$_3$)$_2$O$_2$Cl, but its magnetic behavior is remarkably different. 

\begin{figure}
\includegraphics[width=\columnwidth]{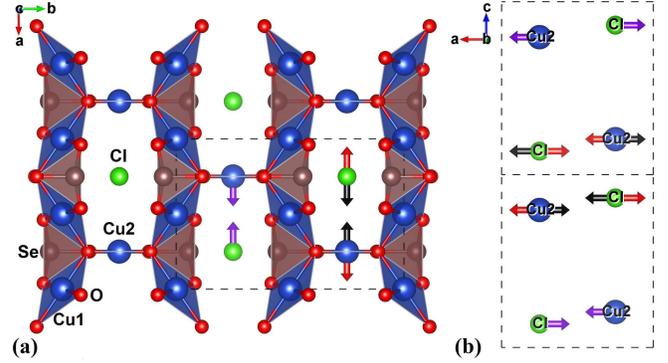}
\caption{\label{fig:structure}
(Color online) Left panel: Single kagome-like layer in the Cu$_3$Bi(SeO$_3$)$_2$O$_2$Cl structure. Bi atoms are omitted for clarity. Right panel: atomic displacements in the adjacent layers upon the structural distortion. The red and black arrows are the displacements associated with the $\Gamma$- and $Z$-phonons, respectively. The purple arrows are the displacements that take place for both $\Gamma$- and Z-phonons.
}
\end{figure}

Crystal structure of Cu$_3$Bi(SeO$_3$)$_2$O$_2$X (X = \mbox{Cl, Br, I}) can be represented as a set of layers stacked along the $c$ direction. Each layer consists of CuO$_{4}$ plaquettes and SeO$_{3}$ trigonal pyramids. Spin-$\frac{1}{2}$ Cu$^{2+}$ ions form a distorted kagome lattice with two non-equivalent nearest-neighbor exchange bonds, both ferromagnetic (FM)~\cite{ioan2015}. These layers are connected by Bi--O bonds, resulting in a net-like structure with halogen atoms residing inside hexagonal tunnels (Fig.~\ref{fig:structure}). According to Millet \mbox{\textit{et al.}}~\cite{millet2001}, the room-temperature crystal structure features orthorhombic $Pmmn$ symmetry with lattice parameters listed in Table~\ref{tab:lattice}.

Both Cl compound (natural francisite) and its Br-containing synthetic analogue are magnetically ordered below $T_N\simeq 24$\,K~\cite{millet2001}. Neutron diffraction studies on single crystals of the Br compound reported $T_N\simeq 27$\,K and revealed canted spin configuration in the $ab$ plane~\cite{pregelj2012}. Net moments of individual layers are canceled macroscopically because of the AFM interlayer coupling. External magnetic field overcomes this coupling and triggers a metamagnetic transition for the field applied along the $c$ direction~\cite{pregelj2012}. The Cl compound reveals a very similar behavior, albeit with a smaller magnetization above the metamagnetic transition: 0.83\,$\mu_B$/Cu for X = Br~\cite{pregelj2012} vs. 0.65\,$\mu_B$/Cu for X = Cl~\cite{miller2012}.

\begin{table}
\caption{\label{tab:lattice}
Experimental lattice parameters $a$, $b$, and $c$ (in\,\r A) of Cu$_3$Bi(SeO$_3$)$_2$O$_2$X at room temperature~\cite{millet2001}.
}
\begin{ruledtabular}
\begin{tabular}{cccc}
       & $a$   & $b$   & $c$   \\\hline
X = Cl & 6.354 & 9.635 & 7.233 \\ 
X = Br & 6.390 & 9.694 & 7.287 \\
X = I  & 6.436 & 9.751 & 7.377 \\
\end{tabular}
\end{ruledtabular}
\end{table}
While a few theoretical studies of this interesting magnetic behavior were reported recently~\cite{ioan2015,nikolaev2016}, and a peculiar field-dependent microwave absorption over several decades in frequency was observed experimentally~\cite{pregelj2015,zorko2016}, one crucial aspect of kagome francisites remains unresolved. Millet~\textit{et al.}~\cite{millet2001} speculated on the presence of a structural phase transition around $100-150$\,K for both Cl and Br compounds, and Miller~\textit{et al.}~\cite{miller2012} indeed observed additional vibration frequencies below 115\,K in the Cl compound, but they were unable to detect any clear signatures of this putative transition in a diffraction experiment. Very recently, Gnezdilov~\textit{et al.}~\cite{gnezdilov2016} claimed that only the Cl compound undergoes a low-temperature structural phase transition, and speculated that the low-temperature phase is polar and even ferroelectric.

In the following, we report a combined computational and experimental study that sheds light on the heavily speculated nature of the structural distortion in kagome francisites. We uncover the distortion in the Cl compound below 110\,K, elucidate its origin, and establish the non-polar, antiferroelectric nature of the low-temperature structure at odds with all earlier proposals~\cite{miller2012,gnezdilov2016}. We show that the Br compound is on the verge of a similar transformation. We also derive the magnetic model for the distorted low-temperature crystal structure and elucidate the differences in the magnetic response of the Cl and Br compounds.

\section{Methods}
Density-functional (DFT) calculations were performed in the \texttt{VASP} pseudo-potential code utilizing the projector-augmented wave (PAW) method~\cite{vasp1,*vasp2}. Perdew-Burke-Ernzerhof flavor of the exchange-correlation potential~\cite{pbe96} corresponding to the generalized gradient approximation (GGA) was used. We performed GGA+$U$ calculations with the on-site Coulomb repulsion $U_{d}$ = 9.5\,eV and Hund's coupling $J_{d}$ = 1.0\,eV in order to account for strong electronic correlations in the Cu $3d$ shell. These parameters are chosen in a semi-empirical manner as providing best agreement with the experimental exchange couplings in previous DFT calculations for francisites~\cite{ioan2015} and for other Cu$^{2+}$ oxide materials~\cite{nath2015,nath2013}. 

The supercell doubled along the $c$-axis and containing 60 atoms was used to account for AFM inter-layer interactions. All atomic positions were fully relaxed with the 1\,meV/\r{A}  convergence criteria for forces and 0.1\,meV for total energy. For electronic structure calculations we used the 500\,eV plane-wave energy cut-off and $2\times 2\times 2$ $k$-mesh centered at the $\Gamma$ point. To check for possible convergence errors, we performed electronic and phonon calculations with the increased $6\times 4\times 3$ $k$-mesh and ensured that the calculated energies, exchange couplings and frequencies did not change.

Phonon calculations were performed by means of the \texttt{PHONOPY}~\cite{phonopy} code using frozen-phonon method to describe elements of the force-constant matrix. Atomic displacements of 0.01\,\r{A} were used to induce non-zero forces in the $2\times 1\times 2$ supercell containing 120 atoms.

To find atomic configurations with the lowest energy, we fully relax the parent structure and calculate phonon spectrum. If the spectrum demonstrates imaginary frequencies, we search for a lower-energy configuration using the corresponding displacement eigenvector. After atoms of the parent structure are displaced and fully relaxed to their new equilibrium positions, we compare energies of the parent and final structural configurations. If the new structure has lower energy, we apply this methodology again for the new structure, and repeat this process until a stable configuration without imaginary phonon frequencies is found.

Powder samples of Cu$_3$Bi(SeO$_3)_2$O$_2$X (X = Cl, Br) for the experimental studies were obtained by grinding crystals prepared following the procedure described in Ref.~\onlinecite{pregelj2012}. High-resolution diffraction data were collected at the ID22 beamline of the European Synchrotron Radiation Facility using the wavelength of 0.41\,\r A. Powders were placed into thin-wall borosilicate capillaries and spun during the experiment. The signal was measured by 9 Si(111) analyzer detectors. All measurements were performed in a He-flow cryostat. \texttt{Jana2006} program~\cite{jana2006} was used for structure refinement.

Magnetic susceptibility of Cu$_3$Bi(SeO$_3)_2$O$_2$X was measured on powders placed into plastic capsules. The measurements were performed using the MPMS 3 SQUID magnetometer from Quantum Design.

\section{Results}
\subsection{Structural instabilities}
\label{sec:relaxation}
\begin{figure}
\includegraphics[width=\columnwidth]{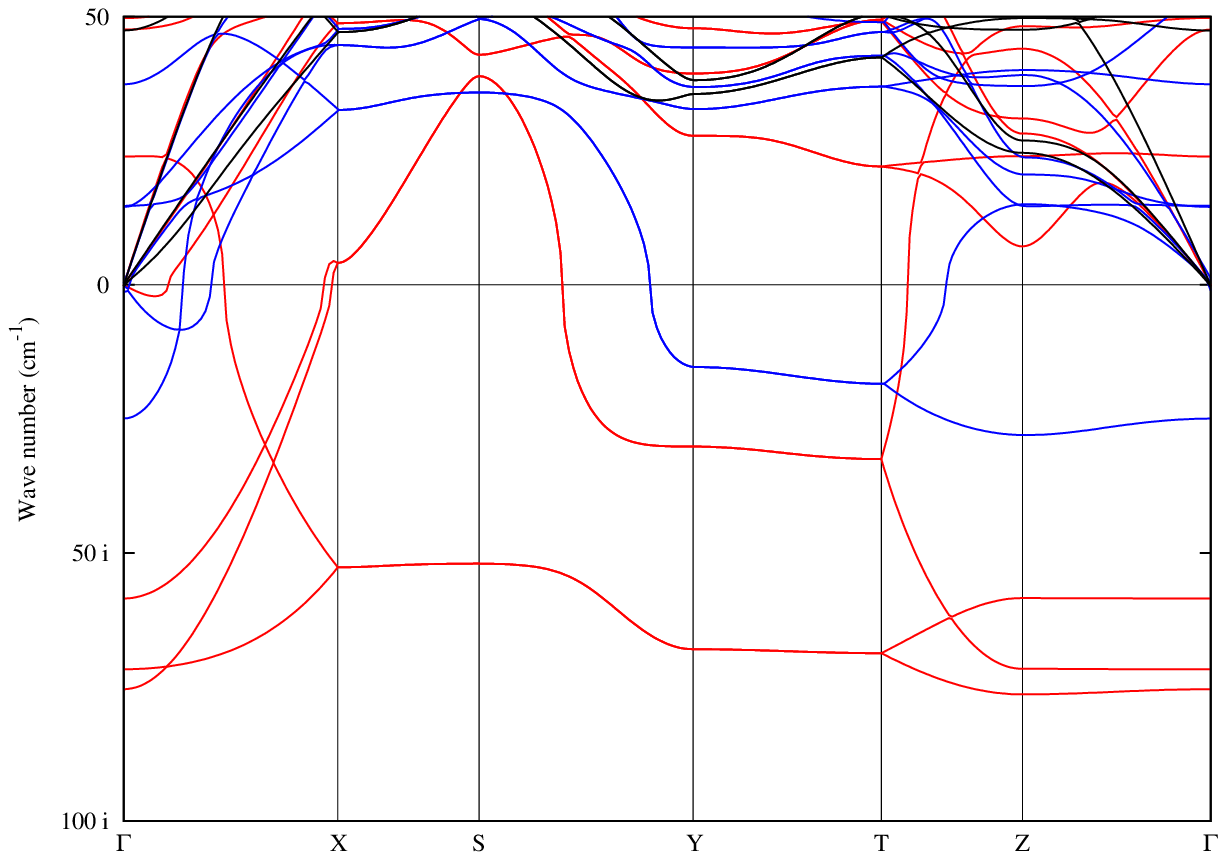}
\caption{\label{fig:Pmmn} (Color online) Phonon spectra for the \textit{Pmmn} structure. The red, blue, and black lines correspond to the structures with Cl, Br, and I atoms, respectively.}
\end{figure}

To check for possible structural instabilities, we calculated phonon spectra of the parent $Pmmn$ structure with X = Cl, Br, and I. This structure was found to be unstable for both X = Cl and Br, but not for I (Fig.~\ref{fig:Pmmn}). We found unstable phonon modes throughout the whole Brillouin zone in the case of the Cl compound and imaginary modes at the $\Gamma(0,0,0)$, $Z(0,0,\frac{1}{2})$, $Y(0,\frac{1}{2},0)$, and $T(0,\frac{1}{2},\frac{1}{2})$ points in the case of Br. 

\begin{table*}
\begin{minipage}{14cm}
\caption{\label{tab:Raman} Computed $\Gamma$-point frequencies of Raman-active modes for the X = Cl compound and their comparison to the experimental 9\,K data from Ref.~\onlinecite{gnezdilov2016}. The modes are split into the $A_g$ and $B_{1g}$ groups according to their symmetries in the $Pmmn$ and $Pcmn$ structures. Asterisks denote vibration modes that were observed only at 100\,K and below.}
\begin{ruledtabular}
\begin{tabular}{cccc @{\hspace{4em}} cccc}
\multicolumn{4}{c}{\textit{$A_{g}$}} & \multicolumn{4}{c}{\textit{$B_{1g}$}} \\
\textit{Pmmn} & \textit{P2$_{1}$mn} & \textit{Pcmn} & \textit{Exp.} &  \textit{Pmmn} & \textit{P2$_{1}$mn} & \textit{Pcmn} & \textit{Exp.}   \\
-& - & 26.6 & 31.1* & -& - & 26.2 & 33.6*  \\
-&  - & 41.5 & 38.3* & 67.4 & 44.1 & 28.8 & 48.6  \\
-& - & - & 57.8* & -& 57.4 & 51.3 & 55.6*  \\
83.2 & 66.3 & 81.8 & 78.2 & 83.2 & 68.6 & 66.8 & 78.1   \\
-& 82.9 & 83.8 & 81.4* & -& 124.5 & 125.1 & 119.7*   \\
90.1 & 87.8 & 90.3 & 86.2& 125.9 & 135.0 & 135.5 & 130.8 \\
-& 136.2 & 137.6 & 137.5* & -& 182.0 & 185.9 & 189.2*  \\
151.4 & 144.3 & 146.2 & 150.2 & -& 251.7 & 258.7 & 266.4* \\
151.4 & 163.7 & 155.3 & 151.6 & -& 293.3 & 295.1 & 297.8*   \\
172.4 & 170.8 & 171.7 & 176.8 & 292.9 & 309.4 & 314.7 & 305.6   \\
-& 189.4 & 195.5 & 198.0* & 392.4 & 394.0 & 391.1 & 397.5   \\
-& 200.6 & 199.6 & 200.4* & 474.2 & 497.5 & 499.3 & 484.7  \\
229.2 & 229.4 & 229.1 & 235.3 &- & 525.1 & 528.0 & 507.1*  \\
-& 304.1 & 305.4 & 322.5* & 703.9 & 698.7 & 701.9  & 723.7    \\
-& 319.6  & 335.9 & 347.5* &- & 705.1 & 706.9 & 730.2*     \\
393.9 & 389.0 & 389.1 & 400.8 &  &  &   &        \\
-& 420.6 & 416.8 & 417.2* & &   &   &     \\
447.2 & 441.8 & 441.1 & 455.9 &  &   &   &   \\
533.9 & 509.9 & 514.2  & 537.5 & &   &   &      \\
-& 547.1 & 548.2  & 555.5*  & &   &   &      \\
-& 566.0 & 564.1  & 586.0* &   &  &   &      \\
-& 666.4 & 668.1  & 689.0* &   &  &  &      \\
752.7 & 753.3 & 752.3  & 774.2 &   &   & &      \\
829.1 & 828.0 & 827.6  & 844.3 &   &   & &      \\
\end{tabular}
\end{ruledtabular}
\end{minipage}
\end{table*}

First, we follow the lowest-lying phonon mode at the $\Gamma$ point and include atomic displacements in our original structure accordingly. Owing to the displacements, the symmetry of the structure changes to \textit{P$2_{1}$mn} (No. 31), and the inversion center is lost. After the relaxation of the $\Gamma$-distorted structure we arrive at the new \textit{P$2_{1}$mn} structure, which is polar. The main difference between the original and relaxed atomic configurations pertains to the positions of the Cl/Br atoms. They move from the $(\frac14,\frac34,z)$ position to the $(\frac14+\Delta,\frac34,z+\delta)$ position with $\Delta=0.49$\,\r{A} and $\delta=0.09$\,\r{A} for the Cl structure and $\Delta=0.14$\,\r{A} and $\delta=0.01$\,\r{A} for the Br structure (see Fig.~\ref{fig:structure}). There are also changes in the positions of Cu and O atoms. The energy difference between the original \textit{Pmmn} and relaxed \textit{P$2_{1}$mn} structures is about 57\,meV/f.u. for Cl and 2\,meV/f.u. for Br. The \textit{P$2_{1}$mn} structure is thus energetically favorable in both cases. 

\begin{table*}
\caption{\label{tab:IR} Infrared-active $\Gamma$-point phonon frequencies computed for the Cl compound in its different structural models ($Pmmn$, $P2_1mn$, and $Pcmn$) and their comparison to the experimental vibration modes from Ref.~\onlinecite{miller2012} (TO frequencies at 7\,K). The vibrations are split into groups depending on their polarization.  Asterisks denote experimental modes that appear below 115\,K.}
\begin{ruledtabular}
\begin{tabular}{cccc @{\hspace{2em}} cccc @{\hspace{2em}} cccc}
\multicolumn{4}{c}{\textit{a}} & \multicolumn{4}{c}{\textit{b}} & \multicolumn{4}{c}{\textit{c}} \\
$Pmmn$ & $P2_{1}mn$ & $Pcmn$ & Exp. & $Pmmn$ & $P2_{1}mn$ & $Pcmn$ & Exp. & $Pmmn$ & $P2_{1}mn$ & $Pcmn$ & Exp. \\
47.6 & - & - & 52.8 & - & 44.1 & 29.1 & 36.3 & - & - & 64.6 & 53.2 \\
 - & 66.3 & 67.7 & 69.9* & 67.1 & 68.6 & 65.8 & 68.3 & 105.4 & 96.2 & 98.4 & 99.8 \\
89.0 & 87.8 & 92.3 & 89.0 & - & - & 100.6 & 99.8* & - & 127.1 & 125.4 & 115.1* \\
- & 90.6 & 101.4 & 101.1* & 114.1 & 113.7 & 113.6 & 115.2 & 138.4 & 146.8 & 136.5 & 144.8 \\
137.0 & 136.2 & 143.3 & 137.6 & - & 124.5 & 127.5 & 128.9* & 159.0 & 158.3 & 165.1 & 161.5 \\
161.7 & 163.7 & 155.5 & 161.9 & 134.4 & 135.0 & 134.3 & 133.5 & 200.4 & 200.0 & 201.2 & 204.0 \\
- & 170.8 & 158.3 & 172.3* & 184.9 & 182.0 & 181.8 & 185.8 & - & 268.5 & - & 273.8* \\
190.0 & 189.4 & - & 191.6 & 254.4 & 251.7 & 249.6 & 256.9 & 295.7 & 289.9 & 302.8 & 284.4 \\
198.9 & 200.6 & 199.3 & 202.1 & - & 288.1 & 267.4 & 276.1* & 322.2 & 322.6 & 320.9 & 337.6 \\
- & 304.1 & 294.7 & 320.0* & 291.6 & 293.3 & 292.1 & 300.3 & 427.1 & 420.5 & 420.2 & 433.5 \\
319.0 & 319.6 & 322.2 & 331.1 & 312.7 & 309.4 & 308.4 & 313.9 & 532.6 & 531.3 & 515.0 & 528.4 \\
 421.9& 420.6 & 419.3 & 422.9 & 450.9 & 448.8 & 448.8 & 456.3 & 583.5 & 572.4 & 566.3 & 554.4 \\
- & 441.8 & 497.3 & 470.2 & - & 472.2 & 472.3 & 484.7* & 769.4 & 769.9 & 782.7 & 794.9 \\
- & 509.9 & 508.3 & 542.4* & 503.2 & 497.5 & 499.4 & 507.0 &  &  &  &   \\
537.8 & 547.1 & 531.3 & 557.3 & 526.7 & 525.1 & 525.1 & 542.3 &  &  &  &   \\
- & 566.0 & 573.6 & 587.2* & - & - & - & 571.4* &  &  &  &   \\
665.4 & 666.4 & 681.2 & 688.2 & - & 705.1 & 702.1 & 716.1* &  &  &  &   \\
- & - & - & 703.8* & - & - & - & 730.0 &  &  &  &   \\
- & 753.3 & 768.2 & 737.0* & 802.2 & 803.0 & 802.2 & 811.3 &  &  &  &   \\
 &  &  &   & - & - & - & 825.0 &  &  &  &   \\
\end{tabular}
\end{ruledtabular}
\end{table*}

A similar analysis for the $Z$-distorted structure results in the centrosymmetric and thus non-polar \textit{Pcmn} space group (No. 62). This distortion results in the doubling of the unit cell along the $c$-axis, because the Cl/Br atoms in the adjacent layers move in opposite directions (Fig.~\ref{fig:structure}, right). After relaxing the distorted structure, we can see that, once again, the Cl/Br atoms reveal the largest displacement amplitude. In comparison to the \textit{P$2_{1}$mn} structure, the \textit{Pcmn} one features $\Delta = 0.42$\,\r{A} and $\delta = 0.05$\,\r{A} for Cl, and $\Delta = 0.17$\,\r{A} and $\delta = 0.01$\,\r{A} for Br. Energy difference between the \textit{Pmmn} and \textit{Pcmn} structures favors the latter by 60\,meV/f.u. and 3\,meV/f.u. for the Cl and Br compounds, accordingly.

Our phonon analysis reveals that the \textit{Pcmn} structure is lowest in energy for both Cl and Br compounds. The \textit{Pcmn} structure is lower in energy than the \textit{P$2_{1}$mn} structure by 3\,meV/f.u. for Cl and by 1\,meV/f.u. for Br~\footnote{Note that we use the non-standard settings $P2_1mn$ ($Pmn2_1$) and $Pcmn$ ($Pnma$) for the sake of comparison with the $Pmmn$ structure.}. 

Analysis of interatomic distances in the relaxed structures suggests that the local environment of Bi, Se, and Cu1 is nearly unchanged upon the distortion. The main changes are related to the mutual positions of the Cu2 and halogen atoms. In the parent $Pmmn$ structure, the halogens are weakly bonded to Cu2 with the \mbox{Cu2--X} distances of 3.205\,\r A for X = Cl and 3.215\,\r A for X = Br~\cite{millet2001}. These distances are determined by the size of the six-fold cavity, which is centered by the halogen atom. Both $P2_1mn$ and $Pcmn$ distorted structures feature the X atoms displaced toward Cu2 in such a way that one \mbox{Cu2--X} distance is shortened to 2.59\,\r A and 2.93\,\r A, while the other distance increases to 3.83\,\r A and 3.51\,\r A for X = Cl and Br, respectively. These shortened Cu2--X distances are clearly correlated with the ionic radius, which is smaller for Cl$^-$ (1.81\,\r A) and larger for Br$^-$ (1.96\,\r A)~\cite{shannon1976}. The absence of the distortion in the X = I compound can be naturally explained by the even larger ionic radius of I$^-$ (2.20\,\r A).

\subsection{$\Gamma$-point phonons}
To facilitate a comparison between our distorted structures and the experimental vibration frequencies~\cite{miller2012,gnezdilov2016}, we calculated $\Gamma$-point phonons. Group-theory analysis suggests the following distribution of optical modes for different crystal symmetries relevant to francisites:
\begin{align*}
 \Gamma^{\rm\,optical}_{Pmmn} &=14B_{1u}^{\rm (IR)}+14B_{2u}^{\rm (IR)}+11B_{3u}^{\rm (IR)} \\
&+12A_{g}^{\rm (R)}+6B_{1g}^{\rm (R)}+9B_{2g}^{\rm (R)}+12B_{3g}^{\rm (R)}+9A_{u},
\end{align*}
\begin{align*}
 \Gamma^{\rm\,optical}_{P2_{1}mn} &=23A_{1}^{\rm (IR)(R)}+20B_{1}^{\rm (IR)(R)}+23B_{2}^{\rm (IR)(R)}\\
&+21A_{2}^{(R)},
\end{align*}
\begin{align*}
 \Gamma^{\rm\,optical}_{Pcmn} &=23B_{1u}^{\rm (IR)}+20B_{2u}^{\rm (IR)}+23B_{3u}^{\rm (IR)} \\
&+24A_{g}^{\rm (R)}+21B_{1g}^{\rm (R)}+24B_{2g}^{\rm (R)}+21B_{3g}^{\rm (R)}+21A_{u},
\end{align*}
where superscripts (IR) and (R) denote infrared-active and Raman-active modes, respectively. All of the above modes distributions were obtained using the \texttt{SMODES}~\cite{smodes} program. 

We discuss Raman modes first. All Raman modes observed at room temperature can be ascribed to the $Pmmn$ structure (Table~\ref{tab:Raman}). Below 120\,K, 21 additional modes were observed and assigned to the loss of inversion symmetry~\cite{gnezdilov2016}, because most of these modes also appeared in the infrared spectra~\cite{miller2012}. We show, however, that the majority of these modes can be well understood in both non-centrosymmetric $P2_1mn$ and centrosymmetric $Pcmn$ structures. The $Pcmn$ symmetry is in fact even more favorable, because it produces low-energy Raman modes around 30\,cm$^{-1}$, which are missing in $P2_1mn$. We do not find any mode of the $A_g$ symmetry around 57.8\,cm$^{-1}$, but the very identification of this mode is somewhat uncertain, because it nearly overlaps with the $B_{1g}$ mode at 55.6\,cm$^{-1}$. Moreover, a Raman-active $B_{3g}$ mode is expected at 60.1\,cm$^{-1}$ in $Pcmn$, and, according to Ref.~\onlinecite{gnezdilov2016}, the $B_{2g}$ and $B_{3g}$ modes may be present in the experimental spectra because of the wide aperture of the collecting spectrometer optics.

The assignment of the infrared modes (Table~\ref{tab:IR}) is generally similar. Most of the modes observed at room temperature are expected in the $Pmmn$ structure, while most of the modes appearing below 115\,K could belong to either $P2_1mn$ or $Pcmn$ symmetries. There are, however, a few experimental modes that could not be correlated with the computed phonons for any of the space groups that we considered. The origin of these modes requires further investigation. They might indicate a more complex structural distortion, but this scenario is excluded by our XRD data (see Sec.~\ref{sec:xrd}). It is worth noting that the infrared modes are obtained by fitting relatively broad peaks in the experimental reflectance spectra~\cite{miller2012} and, therefore, their frequencies are less certain than the frequencies of the Raman modes, where all but one vibrations could be identified uniquely (Table~\ref{tab:Raman}). 

We conclude that additional vibration modes observed in the X = Cl compound below 115\,K indicate deviations from the room-temperature $Pmmn$ symmetry, but the type of the distortion can not be determined unambiguously. Both $P2_1mn$ and $Pcmn$ structures are generally consistent with the experimental lattice vibrations, so either of these structures or even their combination could be formed upon the 115\,K transition. In the following, we probe the low-temperature crystal structure of the X = Cl compound directly and demonstrate its predominantly non-polar, antiferroelectric nature in agreement with our DFT results presented in Sec.~\ref{sec:relaxation}.

\begin{figure}
\includegraphics{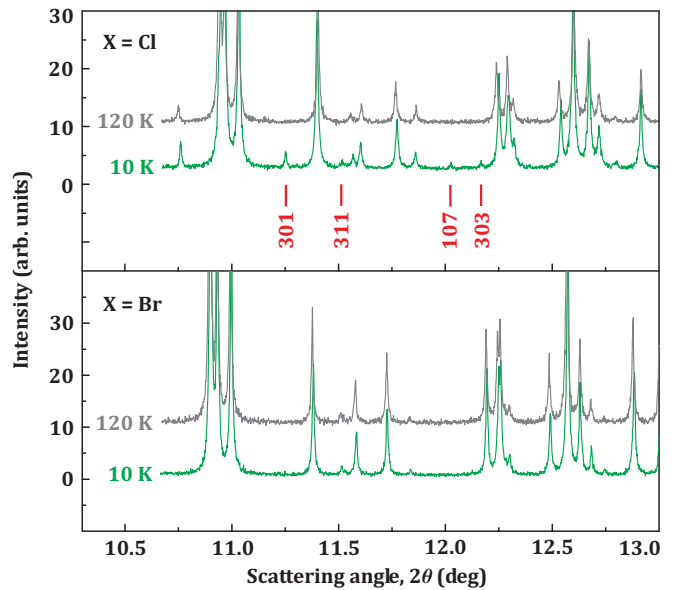}
\caption{\label{fig:xrd}
(Color online) High-resolution synchrotron XRD data for Cu$_3$Bi(SeO$_3)_2$O$_2$X (X = Cl, Br) showing the superstructure reflections at 10\,K for X = Cl (top) and lack thereof for X = Br (bottom). The superstructure reflections are indexed on the $a\times b\times 2c$ supercell. 
}
\end{figure}
\subsection{Low-temperature crystal structure and magnetic behavior}
\label{sec:xrd}
Low-temperature synchrotron XRD data were collected for both Cl and Br-containing compounds, where structural distortions were expected from DFT. The Cl compound will be discussed first. All peaks on its 200\,K pattern could be indexed assuming the unit cell of the undistorted $Pmmn$ structure. Below 100\,K, several weak reflections incompatible with this structure appeared (Fig.~\ref{fig:xrd}). They could be indexed in the unit cell with the doubled $c$-parameter in agreement with our expectations. Indeed, the low-temperature crystal structure of Cu$_3$Bi(SeO$_3)_2$O$_2$Cl was successfully refined in the space group $Pcmn$ revealing the anticipated displacements of Cl atoms to the $(\frac14+\Delta_{\rm Cl},\frac34,z_{\rm Cl})$ position. The Cu2 atoms are also displaced to the $(\frac14+\Delta_{\rm Cu2},\frac34,z_{\rm Cu2})$ position resulting in the shortening of the Cu--Cl2 distance from 3.205\,\r A in the undistorted structure to 2.782(6)\,\r A in the distorted structure (Fig.~\ref{fig:structure}).

Our calculations suggest that the non-polar $Pcmn$ and polar $P2_1mn$ structures have similar energies. The simultaneous formation of these two structures will produce a disordered atomic arrangement with the doubled $c$ parameter and the Cl/Cu2 atoms occupying both $(\frac14+\Delta_i,\frac34,z_i)$ and $(\frac14-\Delta_i,\frac34,z_i)$ positions of the $Pcmn$ space group, where $i$ stands for Cl and Cu2 (these positions are labeled as, e.g., Cl and Cl' in Table~\ref{tab:structure}). To explore this possibility, we introduced Cl and Cu2 atoms into the $(\frac14-\Delta_i,\frac14,z_i)$ positions with the same fractional occupancy $f_-$, whereas the occupancy of the $(\frac14+\Delta_i,\frac14,z_i)$ positions was set to $f_+$. The $f_+$ and $f_-$ parameters are thus equivalent to phase fractions of the $Pcmn$ and $P2_1mn$ structures. Structure refinement under the constraint $f_++f_-=1$ resulted in $f_+=0.90(2)$ and $f_-=0.10(2)$ at 10\,K indicating that Cu$_3$Bi(SeO$_3)_2$O$_2$Cl is predominantly in the non-polar $Pcmn$ structure, while the non-zero value of $f_-$ is due to trace amounts of the polar $P2_1mn$ structure or due to residual disorder of the Cl position in the $Pcmn$ structure. 

\begin{figure}
\includegraphics{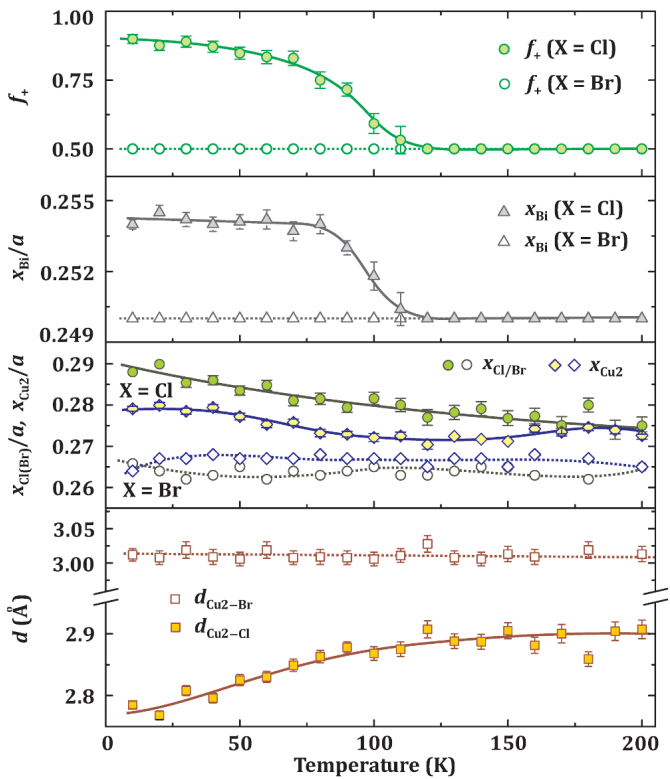}
\caption{\label{fig:refinement}
(Color online) Structure refinement for Cu$_3$Bi(SeO$_3)_2$O$_2$X (X = Cl, Br) at different temperatures. From top to bottom: i) the occupancy $f_+$ of the $(\frac14+\Delta,\frac34,z)$ position of Cl/Br and Cu2; ii) the $x$-coordinate of Bi atoms; iii) the $x$-coordinates of the Cu2 and Cl/Br atoms; iv) the Cu2--Cl/Br distance. The values of $f_+$ and $x_{\rm Bi}$ were fixed to 0.5 and 0.25, respectively, above 120\,K for the Cl compound and at all temperatures for the Br compound. The open and filled symbols are for X = Br and Cl, respectively. The error bars are from the Rietveld refinement. The solid and dashed lines are guides-for-the-eye only.
}
\end{figure}

Temperature dependence of $f_+$ (Fig.~\ref{fig:refinement}, top) supports the residual disorder scenario. If both $Pcmn$ and $P2_1mn$ phases were formed below 115\,K, one would expect an abrupt change of $f_+$ from 0.50 (complete disorder) to its low-temperature value $f_+=0.90(2)$. Instead, we observe a gradual increase in $f_+$ suggesting the slow formation of the $Pcmn$ order upon cooling. The temperature evolution of $f_+$ is reminiscent of a second-order phase transition. Likewise, weak displacements of Bi atoms (Fig.~\ref{fig:refinement}) set in below 115\,K and saturate at low temperatures, as expected for a second-order phase transition. On the other hand, no clear transition anomaly can be seen in the positions of the strongly displaced Cu2 and Cl atoms. The displacement of the Cl atoms ($\Delta_{\rm Cl}$) is slowly decreasing upon heating, but above 130\,K this effect is balanced by the enhanced displacement of Cu2. Therefore, the Cu2--Cl distance increases from 2.78\,\r A at 10\,K to about 2.88\,\r A at 120\,K and stays constant upon further heating (Fig.~\ref{fig:refinement}, bottom).
\begin{table}
\caption{\label{tab:structure}
Refined atomic positions for Cu$_3$Bi(SeO$_3)_2$O$_2$Cl at 10\,K (upper lines) and 200\,K (bottom lines)~\cite{supplement}. The $f_i$ are occupancy factors, and $U_{\rm iso}$ are atomic displacement parameters (ADP) in 10$^{-2}$\,\r A$^2$. The space group is $Pcmn$ (No. 62). The lattice parameters are $a=6.35043(3)$\,\r A, $b=9.62715(4)$\,\r A, and $c=14.42617(6)$\,\r A at 10\,K and $a=6.34983(3)$\,\r A, $b=9.62987(4)$\,\r A, and $c=14.45590(7)$\,\r A at 200\,K. Refinement residuals are $R_I=0.021$, $R_p=0.081$, and $R_{wp}=0.107$ at 10\,K and $R_I=0.026$, $R_p=0.104$, and $R_{wp}=0.138$ at 200\,K. The splitting of the Cu2 and Cl positions is explained in the text. 
}
\begin{ruledtabular}
\begin{tabular}{cccccc}
 Atom & $x/a$ & $y/b$ & $z/c$ & $f_i$ & $U_{\rm iso}$ \\
 Bi   & 0.2540(2) & $\frac14$ & 0.87077(4) & 1.0 & 0.30(1) \\\smallskip
      & $\frac14$ & $\frac14$ & 0.87059(6) & 1.0 & 0.76(2) \\
 Cu1  & 0.9982(10) & $-0.0014(4)$ & 0.7509(3) & 1.0 & 0.53(3) \\\smallskip
      & 0.9969(12) & 0.0033(7)    & 0.7495(8) & 1.0 & 1.23(4) \\
 Cu2  & 0.2791(6)  & $\frac14$    & 0.1466(1) & 0.90(2) & 0.34(7) \\\smallskip
      & 0.2726(9)  & $\frac14$    & 0.1462(1) & 0.50    & 0.58(9) \\
 Cu2' & 0.2209(6)  & $\frac14$    & 0.1466(1) & 0.10(2) & 0.34(7) \\\smallskip
      & 0.2274(9)  & $\frac14$    & 0.1462(1) & 0.50    & 0.58(9) \\
 Se   & 0.2557(4)  & 0.5570(1)    & 0.0546(1) & 1.0     & 0.44(2) \\\smallskip
      & 0.2507(10) & 0.5567(1)    & 0.0548(1) & 1.0     & 1.06(3) \\
 O1   & 0.259(3)   & 0.1112(5)    & 0.2439(4) & 1.0     & 0.09(8) \\\smallskip
      & 0.233(3)   & 0.1158(7)    & 0.2448(5) & 1.0     & 0.5(1)  \\
 O2   & 0.032(2)   & 0.5728(10)   & 0.1216(8) & 1.0     & 0.09(8) \\\smallskip
      & 0.047(2)   & 0.594(1)     & 0.127(2)  & 1.0     & 0.5(1)  \\
 O3   & 0.049(1)   & 0.593(1)     & 0.6309(8) & 1.0     & 0.09(8) \\\smallskip
      & 0.023(2)   & 0.570(1)     & 0.628(1)  & 1.0     & 0.5(1)  \\
 O4   & 0.273(1)   & 0.1157(6)    & 0.0454(4) & 1.0     & 0.09(8) \\\smallskip
      & 0.275(2)   & 0.1170(7)    & 0.0441(5) & 1.0     & 0.5(1)  \\
 Cl   & 0.2880(8)  & $\frac34$    & 0.8225(2) & 0.90(2) & 0.18(11) \\\smallskip
      & 0.275(2)   & $\frac34$    & 0.8229(3) & 0.50    & 2.2(2)   \\
 Cl'  & 0.2120(2)  & $\frac34$    & 0.8225(2) & 0.10(2) & 0.18(11) \\
      & 0.225(2)   & $\frac34$    & 0.8225(2) & 0.50    & 2.2(2)   \\
\end{tabular}
\end{ruledtabular}
\end{table}

All these observations suggest that the displacements of Cl and Cu2 (Fig.~\ref{fig:structure}) are present in both low-temperature and high-temperature crystal structures of Cu$_3$Bi(SeO$_3)_2$O$_2$Cl. The shortening of the Cu2--Cl distance is energetically highly favorable and results in the energy gain of about 60\,meV/f.u. (see Sec.~\ref{sec:relaxation}) that exceeds the energy of thermal fluctuations within the temperature range of our study. Therefore, the shortened Cu2--Cl distances are formed already at high temperatures. In fact, we could even make a better structure refinement at 200\,K using the disordered $Pcmn$ model instead of the original $Pmmn$ model~\footnote{At 200\,K, refinement residuals are 0.035 in $Pmmn$ and 0.026 in $Pcmn$ for 20 and 30 refinable parameters, respectively.}. Such a behavior is very common for order-disorder transitions, where local distortions typically persist above the transition temperature~\cite{teslic1998,tsirlin2010}.

\begin{figure}
\includegraphics{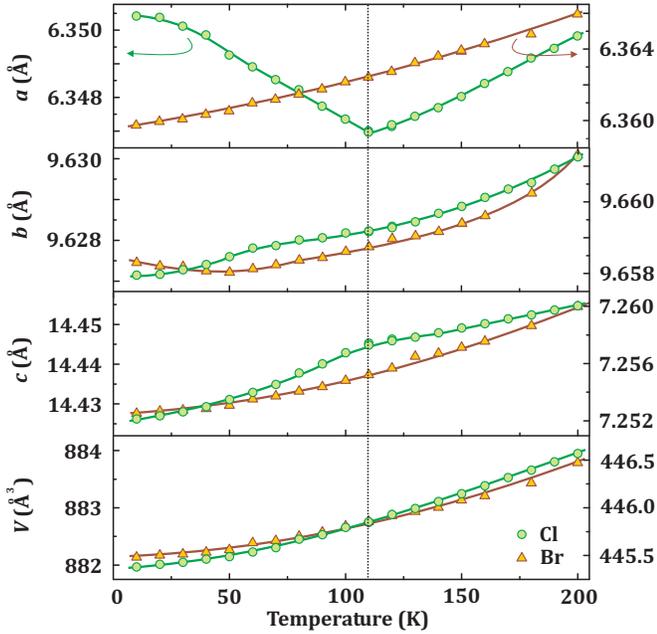}
\caption{\label{fig:expansion}
(Color online) Temperature evolution of lattice parameters ($a,b,c$) and unit cell volume ($V$) for Cu$_3$Bi(SeO$_3)_2$O$_2$X (X = Cl and Br). The dotted line denotes the transition temperature of 110\,K in the Cl compound. The error bars are smaller than the symbol size. The solid lines are guide-for-the-eye only.
}
\end{figure}
The structural phase transition in Cu$_3$Bi(SeO$_3)_2$O$_2$Cl manifests itself by anomalies in thermal expansion (Fig.~\ref{fig:expansion}). The most prominent effect is seen for the $a$ parameter that decreases upon heating to 110\,K and increases upon further heating. On the other hand, there is only weak effect in the temperature evolution of the unit cell volume. No abrupt change of the cell volume is observed around 110\,K, indicating the second-order nature and, thus, order-disorder type of the transition. The transition temperature of about 110\,K (see Figs.~\ref{fig:refinement} and \ref{fig:expansion}) is in agreement with Refs.~\onlinecite{miller2012,gnezdilov2016}.

\begin{table}
\caption{\label{tab:structure-Br}
Refined atomic positions for Cu$_3$Bi(SeO$_3)_2$O$_2$Br at 10\,K (upper lines) and 200\,K (bottom lines)~\cite{supplement}. The $U_{\rm iso}$ are atomic displacement parameters (ADP) in 10$^{-2}$\,\r A$^2$. The space group is $Pmmn$ (No. 59). The lattice parameters are $a=6.35974(2)$\,\r A, $b=9.65832(3)$\,\r A, and $c=7.25252(2)$\,\r A at 10\,K and $a=6.36595(2)$\,\r A, $b=9.66039(3)$\,\r A, and $c=7.25998(2)$\,\r A at 200\,K. Refinement residuals are $R_I=0.026$, $R_p=0.088$, and $R_{wp}=0.118$ at 10\,K and $R_I=0.026$, $R_p=0.107$, and $R_{wp}=0.148$ at 200\,K.
}
\begin{ruledtabular}
\begin{tabular}{cccccc}
 Atom & Site & $x/a$ & $y/b$ & $z/c$ & $U_{\rm iso}$ \\
 Bi   & $2a$ & $\frac14$ & $\frac14$ & 0.24052(9) & 0.33(1) \\\smallskip
      &      & $\frac14$ & $\frac14$ & 0.2401(1)  & 0.64(2) \\
 Cu1  & $4c$ & 0         & 0         & 0          & 0.52(3) \\\smallskip
      &      & 0         & 0         & 0          & 1.00(4) \\
 Cu2\footnotemark[1]  & $4f$ & 0.264(1)   & $\frac14$    & 0.7924(2) & 0.28(6) \\\smallskip
      &      & 0.269(1)   & $\frac14$    & 0.7919(3) & 0.55(9) \\
 Se   & $4e$ & $\frac14$  & 0.5564(1)    & 0.6116(1) & 0.39(2) \\\smallskip\footnotetext[1]{Occupancy factor $f_i=\frac12$.}
      &      & $\frac14$  & 0.5566(1)    & 0.6121(2) & 0.71(3) \\
 O1   & $4e$ & $\frac14$  & 0.1099(5)    & 0.9920(8) & 0.05(7) \\\smallskip
      &      & $\frac14$  & 0.1132(7)    & 0.992(1)  & 0.7(1)  \\
 O2   & $8g$ & 0.0385(5)  & 0.5856(4)    & 0.7573(6) & 0.05(7) \\\smallskip
      &      & 0.0390(7)  & 0.5837(5)    & 0.7579(8) & 0.7(1)  \\
 O3   & $4e$ & $\frac14$  & 0.1127(6)    & 0.5924(8) & 0.05(7) \\\smallskip
      &      & $\frac14$  & 0.1145(7)    & 0.592(1)  & 0.7(1)  \\
 Br\footnotemark[1]   & $4f$ & 0.266(1)   & $\frac34$    & 0.1555(2) & 0.16(5) \\
      &      & 0.266(1)   & $\frac34$    & 0.1573(2) & 1.20(7) \\
\end{tabular}
\end{ruledtabular}
\end{table}
The behavior of the X = Br compound is remarkably different. Its thermal expansion is featureless (Fig.~\ref{fig:expansion}) with the exception of a small dip in the temperature dependence of the $a$ parameter around 50\,K. No superstructure reflections were observed down to 10\,K (Fig.~\ref{fig:xrd}, bottom). The crystal structure could be fully refined in the $Pmmn$ space group. However, the splitting of the Cu2 and Br positions revealed non-negligible displacements of these atoms (Table~\ref{tab:structure-Br}) resulting in the Cu2--Br distance of 3.02(1)\,\r A at 10\,K, which is much shorter than 3.21\,\r A in the ideal structure. In contrast to the X = Cl compound, this shortened Cu2--Br distance is temperature-independent within the sensitivity of our measurement (Fig.~\ref{fig:refinement}, bottom). We conclude that the Br compound does not undergo the low-temperature structural transition, but it features local atomic displacements in agreement with the DFT predictions discussed in Sec.~\ref{sec:relaxation}.
\begin{figure}
\includegraphics{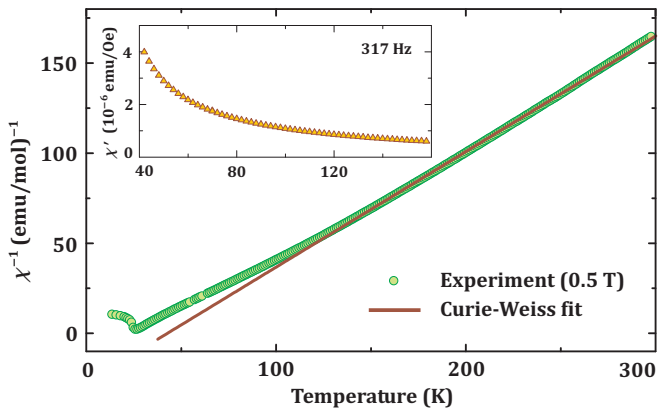}
\caption{\label{fig:chi}
(Color online) Inverse dc magnetic susceptibility of Cu$_3$Bi(SeO$_3)_2$O$_2$Cl measured in the applied field of 0.5\,T (circles) and the Curie-Weiss fit (solid line), as described in the text. The inset shows real part of the ac susceptibility ($\chi'$) measured at the frequency of 317\,Hz with the field amplitude of 5\,Oe.
}
\end{figure}

To further pinpoint the structural phase transition, we performed magnetic susceptibility measurements in both dc and ac regimes. According to Ref.~\onlinecite{millet2001}, real part of the ac susceptibility ($\chi'$) of the Cl compound shows a kink around 120-130\,K that was allegedly interpreted as a signature of a structural phase transition. Our data are different and reveal smooth evolution of $\chi'(T)$ across the transition temperature (Fig.~\ref{fig:chi}). A peculiarity in the inverse dc susceptibility $1/\chi(T)$ around 150\,K~\cite{millet2001,miller2012} was also reported and tentatively ascribed to a structural effect. We observed only a weak and very smooth bending of $1/\chi(T)$, which is most likely indicative of the gradual onset of spin-spin correlations. The bending in our data is much weaker than in the data reported previously~\cite{millet2001,miller2012}. The data in the temperature range between 150\,K and 300\,K follow the Curie-Weiss law $\chi=C/(T+\theta)$ with $C=1.559$\,emu\,K/mol and $\theta=-43$\,K. The resulting effective moment of $\mu_{\rm eff}=2.03$\,$\mu_B$/Cu corresponds to a spin-$\frac12$ ion with an isotropic $g$-value of $g=2.34$.

\subsection{Magnetic model}
In the following, we propose microscopic magnetic model for the experimental low-temperature crystal structure of Cu$_3$Bi(SeO$_3)_2$O$_2$Cl~\footnote{We used the ordered version of the 10\,K crystal structure neglecting residual disorder related to the positions Cu2' and Cl'.}. To this end, we calculate individual exchange couplings from total energies of collinear and non-collinear spin configurations using the method proposed by Xiang \textit{et al.}~\cite{xiang2011}. The resulting exchange couplings $J_i$ are listed in Table~\ref{table:couplings}. We have also calculated $a$-components of Dzyaloshinsky-Moriya (DM) vectors on the bonds $J_{11}'$ and $J_{12}'$, because those DM-components are responsible for stabilizing canted order in the undistorted ($Pmmn$) francisite structure~\cite{ioan2015}.

\begin{figure}
\includegraphics[width=\columnwidth]{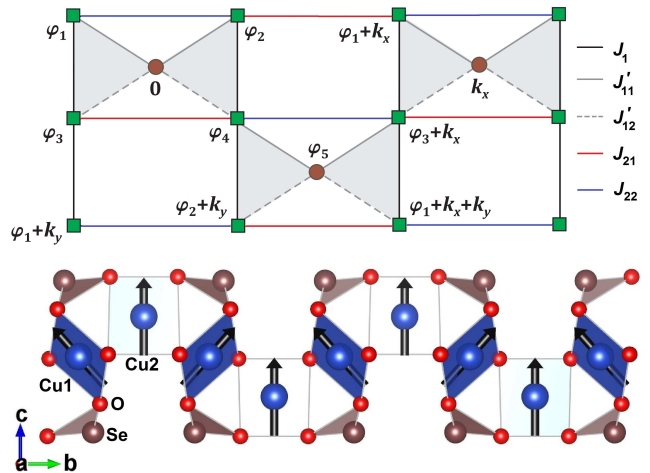}
\caption{\label{fig:ang} (Color online) Top panel: magnetic model for the $Pcmn$ structure showing an arbitrary planar magnetic order defined by five angles $\varphi_1-\varphi_5$ and the components $k_x,k_y$ of the propagation vector. Bottom panel: ground-state spin configuration (canted order) obtained for the exchange parameters from Table~\ref{table:couplings}.}
\end{figure}

Although the unit cell of the low-temperature phase is doubled along the $c$ direction, two layers within one unit cell are related by the glide-plane symmetry and feature same exchange couplings. On the other hand, some of the exchange paths in the $ab$ plane become non-equivalent, and the number of independent exchange parameters increases from three in the room-temperature $Pmmn$ phase ($J_1$, $J_1'$, and $J_2$) to five in the low-temperature $Pcmn$ phase. Namely, $J_1$ splits into $J_{11}'$ and $J_{12}'$, whereas $J_2$ splits into $J_{21}$ and $J_{22}$ (Fig.~\ref{fig:ang}, top). All couplings along the $a$ direction remain equivalent ($J_1$). 

\begin{table}
\caption{\label{table:couplings} Computed magnetic interactions in the low-temperature structure of Cu$_3$Bi(SeO$_3$)$_2$O$_2$Cl: the Cu--Cu distances (in \r{A}), the relevant Cu--O--Cu ($J_1, J_{11}'$, $J_{12}'$, $d_{11a}$ and $d_{12a}$) and Cu--O--O ($J_{21}$ and $J_{22}$) angles $\varphi$ (in deg), exchange couplings $J_i$ and DM-components $d_{ia}$ (in K), as obtained from GGA+$U$ calculations with $U_{d}=9.5$\,eV and $J_{d}=1$\,eV.}
\begin{ruledtabular}
\begin{tabular}{cccr}
Bond type & \textit{$d_{\rm{Cu-Cu}}$} & $\varphi$ & $J_i$ \\
\hline
$J_{1}$ & 3.18 & 112 & $-74$  \\
$J_{11}^{'}$ & 3.31 & 117 & $-36$  \\
$d_{11a}$ &  &  & $19$  \\
$J_{12}^{'}$ & 3.18 & 112 & $-67$  \\
$d_{12a}$ &  & & $13$  \\
$J_{21}$ & 4.84 & 125 & 53  \\
$J_{22}$ & 4.78 & 123 & 45  
\end{tabular}
\end{ruledtabular}
\end{table}

Exchange couplings in Table~\ref{table:couplings} follow peculiarities of the low-temperature $Pcmn$ structure. The coupling $J_{11}'$ is less ferromagnetic than $J_{12}'$, because the corresponding Cu--O--Cu angle is larger (117$^{\circ}$ vs. 112$^{\circ}$), in agreement with Goodenough-Kanamori-Anderson rules. Likewise, the coupling $J_{21}$ is stronger than $J_{22}$, because its Cu--O--O--Cu superexchange pathway is less curved (the \mbox{Cu--O--O} angle is closer to 180$^{\circ}$ expected for the linear pathway). These exchange couplings can be compared to the experimental Curie-Weiss temperature $\theta$ by adding interactions at each Cu site and averaging over the Cu1 and Cu2 sites:
\begin{equation}
 \theta=\frac{1}{3}\left(2\sum_iz_i^{\rm Cu1}J_i+\sum_iz_i^{\rm Cu2}J_i\right),
\end{equation}
where $z_i$ is the number interactions of type $i$ per Cu site. We find $\theta=-42$\,K in excellent agreement with the experimental value of $\theta\simeq -43$\,K.

We will now analyze the magnetic ground state expected on this distorted kagome lattice. We restrict ourselves to planar spin configurations, because the large $a$-component of DM interactions on the bonds $J_1'$ ($J_{11}'$ and $J_{12}'$ in the $Pcmn$ structure), $d_{1a}\simeq 12$\,K~\cite{ioan2015}, puts spins in the $bc$ plane. An arbitrary magnetic ground state is then described by five angles $\varphi_1-\varphi_5$ defining relative spin directions within one repetition unit of the spin lattice (Fig.~\ref{fig:ang}, top) and by two additional parameters $k_x$ and $k_y$ standing for the periodicity of the magnetic structure along the $a$ and $b$ directions, respectively. The ground-state spin configuration is obtained by minimizing the classical energy for the five exchange couplings listed in Table~\ref{table:couplings} and by including $d_{11a}$ and $d_{12a}$, the $a$-components of the DM couplings on the bonds $J_{11}'$ and $J_{12}'$. According to Ref.~\onlinecite{ioan2015}, this DM-components is crucial for removing classical degeneracy and stabilizing canted magnetic order in kagome francisites. 

By minimizing the classical energy of the spin Hamiltonian, we arrive at $\varphi_{1}=\varphi_{3}=-\varphi_{2}=-\varphi_{4}$, $k_{x}=k_{y}=\varphi_{5}=0$, corresponding to the same canted phase as in the room-temperature $Pmmn$ structure (Fig.~\ref{fig:ang}, bottom). The stability of this result was verified by varying $d_{11a}$ and $d_{12a}$. The formation of the proposed canted phase is rooted in the symmetry of the DM vectors having opposite signs on different lattice bonds (Fig.~\ref{fig:ang}, top). The absolute values of $d_{11a}$ and $d_{12a}$ has only marginal effect on the size of the canting angle and do not change the result qualitatively.

Using $d_{11a} = 19$\,K and $d_{12a}=13$\,K, we arrive at the canting angle of $\theta=63.6^{\circ}$ in Cu$_3$Bi(SeO$_3)_2$O$_2$Cl for our model. While this angle has not been measured directly, it can be estimated from the magnetization above the metamagnetic transition when external field is applied along the $c$ direction. Using the experimental value of $M_r=0.65$\,$\mu_B$/Cu~\cite{miller2012}, we find $\theta=67^{\circ}$ in reasonable agreement with our DFT result. Importantly, the structural distortion in the Cl compound is responsible for the reduction in $M_r$ and, consequently, for the increase in $\theta$ compared to the Br compound, where $M_r=0.83$\,$\mu_B$/Cu and $\theta\simeq 51.6^{\circ}$ have been reported~\cite{pregelj2012}. The change of the canting angle can be traced back to the disparity of $J_{11}'$ and $J_{12}'$, and to the reduction in the absolute value of the averaged coupling $\bar{J_1'}=(J_{11}'+J_{12}')/2$ upon the distortion.

\section{Discussion and Summary}
Our results shed light on the interesting and so far poorly understood aspect of kagome francisites, their low-temperature structural distortion that was actively speculated in the earlier literature but never probed in a direct experiment. Using DFT calculations of lattice dynamics, we demonstrate structural instability of francisites with smaller halogen atoms and confirm this instability for the Cl compound by high-resolution XRD. The main effect of the distortion is the collective displacement of Cl and Cu2 resulting in the shortening of the Cu2--Cl distance from 3.21\,\r A in the undistorted $Pmmn$ structure to 2.78\,\r A in the distorted $Pcmn$ structure. It is thus natural that the tangible distortion effect could be observed in the Cl compound, no distortion was envisaged for the I compound, whereas the Br compound is midway between the two. It features local displacements of the Br and Cu2 atoms, but lacks a distinct low-temperature phase where these displacements would be ordered. 

The distortion is rooted in the size of the halogen atom. While Cl is too small for its position in the ideal francisite structure, iodine is big enough to be stable there. It is worth noting that the Cu2 and Cl atoms develop similar displacements and, thus, similar Cu2--Cl distances in both low-temperature and high-temperature crystal structures. In fact, we find that the crystal structure above 115\,K is better described by the disordered $Pcmn$ model than by the high-symmetry $Pmmn$ model considered in the previous literature~\cite{millet2001}. This situation is not uncommon for ferroelectrics and antiferroelectrics, where electric dipoles survive well above the transition temperature $T_c$, and the (anti)ferroelectric transition involves long-range ordering of these dipole species that have been pre-formed at a temperature much higher than $T_c$.

The shortened Cu2--Cl distances of 2.78\,\r A are well in line with crystal structures of other Cu$^{2+}$ minerals containing halogen atoms. For example, haydeeite Cu$_3$Mg(OH)$_6$Cl$_2$ features Cu--Cl distances of 2.76\,\r A~\cite{malcherek2007}, and in herbertsmithite Cu$_3$Zn(OH)$_6$Cl$_2$ the Cu--Cl distance is 2.77\,\r A~\cite{shores2005}. In francisite, this distance is underestimated by DFT that predicts the value of 2.59\,\r A, 0.19\,\r A shorter than in the experiment. A similar underestimate can be seen for the Br compound, where DFT predicts the Cu2--Br distance of 2.93\,\r A, while experimentally the displacements of Cu2 and Br produce the shortest \mbox{Cu2--Br} distance of about 3.0\,\r A. The Br compound does not show any long-range ordering of the Cu2--Br dipoles and thus lacks any structural transformation down to 10\,K. We ascribe this effect to the very small energy differences between the distorted $Pcmn$/$P2_1mn$ structures and the ideal $Pmmn$ structure. They differ in energy by not more than 3\,meV/f.u. Effects like quantum fluctuations, which are important for delicate energy balance in ferroelectrics and antiferroelectrics~\cite{mueller1979,akbarzadeh2004}, may eventually suppress the distortion and stabilize the averaged $Pmmn$ structure. 

Our results are in agreement with the recent report~\cite{gnezdilov2016} that proposed the structural phase transition around 120\,K for the Cl compound and lack thereof for the Br compound. By analyzing the infrared and Raman frequencies, the authors of Ref.~\onlinecite{gnezdilov2016} speculated that the low-temperature phase of Cu$_3$Bi(SeO$_3)_2$O$_2$Cl is ferroelectric ($P2_1mn$ symmetry), but our results refute their conjecture. The observation of the superstructure reflections indicates the doubling of the unit cell, while the structure refinement of the 10\,K data suggests that the sample is predominantly in the non-polar $Pcmn$ phase. In fact, both infrared and Raman frequencies can be interpreted in the framework of the non-polar $Pcmn$ structure without invoking the polar phase. The anomaly of dielectric permittivity~\cite{gnezdilov2016} is reminiscent of an antiferroelectric transition, and no evidence of ferroelectric polarization below 120\,K has been reported.

We conclude that Cu$_3$Bi(SeO$_3)_2$O$_2$Cl features the non-polar $Pcmn$ symmetry at low temperatures, whereas the Br and, presumably, the I compounds retain the $Pmmn$ symmetry of the average crystal structure down to low temperatures. Therefore, kagome francisites are either paraelectric or antiferroelectric, although the polar and potentially ferroelectric $P2_1mn$ phase of the Cl compound is close in energy, only 3\,meV/f.u. away from the lowest-energy antiferroelectric structure. This situation is remarkably similar to the magnetism of these materials, where large net magnetization produced by the canted magnetic order in the $ab$ plane is suppressed by the antiferromagnetic interlayer coupling. Magnetic field of less than 1\,T applied along the $c$ direction renders francisites ferrimagnetic~\cite{pregelj2012,miller2012}. A similar strategy might be pursued for dielectric properties, and further investigation of this aspect is highly desirable.

\begin{acknowledgments}
AT is grateful to Oksana Zaharko, Ioannis Rousochatzakis, and Johannes Richter for fruitful discussions. We acknowledge Carlotta Giacobbe for her work as the ESRF local contact and Valery Verchenko and Darya Nasonova for their support during the synchrotron measurement. The provision of the ESRF beamtime at ID22 is kindly acknowledged. This study was supported by the Supercomputing Center of Lomonosov Moscow State University~\cite{msu}. The work of DP and VM was supported by the grant program of the Russian Science Foundation 14-12-00306. AT was supported by Federal Ministry for Education and Research through the Sofja Kovalevskaya Award of Alexander von Humboldt Foundation. The work in Augsburg was partly supported by the DFG via the Transregional Research Collaboration TRR 80: From Electronic Correlations to Functionality (Augsburg/Munich/Stuttgart) and SNF SCOPES project IZ73Z0 152734/1. AJ was supported by the Deutsche Forschungsgemeinschaft (DFG, German Research Foundation) -- JE 748/1.
\end{acknowledgments}

%

\end{document}